\documentclass[pra,aps,twocolumn]{revtex4}
\usepackage{graphicx}
\usepackage{amsbsy}
\usepackage{amsmath,amsthm}
\usepackage{amsfonts}

\newcommand{\ket}[1]{|{#1}\rangle}
\newcommand{\nn}{\nonumber}
\newtheorem{thm}{Theorem}
\newtheorem{cor}{Corollary}

\begin{document}

\title{Qubit channels that achieve capacity with two states}

\author{Dominic W.\ Berry}
\affiliation{Department of Physics, The University of Queensland, Brisbane,
Queensland 4072, Australia}

\begin{abstract}
This paper considers a class of qubit channels for which three states are
always sufficient to achieve the Holevo capacity. For these channels it is
known that there are cases where two orthogonal states are sufficient, two
non-orthogonal states are required, or three states are necessary. Here a
systematic theory is given which provides criteria to distinguish cases where
two states are sufficient, and determine whether these two states should be
orthogonal or non-orthogonal. In addition, we prove a theorem on the form of the
optimal ensemble when three states are required, and present efficient methods
of calculating the Holevo capacity.
\end{abstract}

\maketitle

\section{Introduction}
A quantum channel is a completely positive and trace preserving (CPTP) map on
quantum states. The condition that it is completely positive means that the
result of the map is a positive operator, and therefore may represent the state
of a system, even if the map acts on one part of an entangled system. The
condition that it is trace preserving ensures that the final state is
normalised. In contrast to unitary operations, quantum channels can increase the
entropy of a state. A quantum channel arises if an ancilla space is added, a
unitary operation is performed between the system and the ancilla, then the
ancilla is traced over to obtain the reduced density operator for the system.

Quantum channels are used to model communication channels, and therefore an
important quantity to consider for these channels is the amount of classical
communication that may be performed. This is often quantified by the Holevo
capacity. The Holevo capacity of a quantum channel $\Phi$ is given by
\begin{equation}
C(\Phi) = \sup_{p_i,\rho_i} S[\Phi(\bar\rho)]-\sum_i p_i S[\Phi(\rho_i)],
\end{equation}
where $\bar\rho=\sum_i p_i \rho_i$, and $S(\sigma)=-{\rm Tr}\sigma\log_2\sigma$
is the von Neumann entropy. The $p_i$ are probabilities, and therefore must be
non-negative and sum to 1. The Holevo capacity is the asymptotic classical
communication that may be achieved using joint measurements on output states,
but unentangled inputs \cite{holevo,schuwes}. In general determining the Holevo
capacity of a channel is a nontrivial task. For the class of channels considered
here, it will be shown that the capacity may be determined in a straightforward
way.

An important issue is the number of states $\rho_i$ that must be considered in
the maximisation. It is well known that, for quantum channels that act upon a
Hilbert space of dimension $d$, the number of states in the ensemble need not
exceed $d^2$ \cite{davies}. In particular, for a qubit channel no more than four
states are required. For the very simple case of unital qubit channels, where
$\Phi(\openone)=\openone$, the capacity is achieved for two orthogonal input
states \cite{kingrus}. For more general qubit channels, the capacity may be
achieved for two non-orthogonal inputs \cite{fuchs}, three states
\cite{kingnath}, or four states may be required \cite{hayashi}.

With the exception of the channels considered in Ref.\ \cite{hayashi}, these
results are all for a class of channels that can require at most three states.
Here we give simple criteria for these channels that, when satisfied, mean that
two states are sufficient. These criteria are not satisfied by the channels
that require three states given in \cite{kingnath}, but are satisfied by
examples given in Refs.\ \cite{kingrus,fuchs,kingnath,cortese} where two states
are sufficient. In addition, we give criteria to determine when the input states
should be orthogonal or non-orthogonal.

This paper is organised as follows. We present the proof of the criteria in
Sec.\ \ref{sec:2st}. Then, in Sec.\ \ref{sec:apply} we give applications of the
result to results presented in previous work. We consider the form of the
optimal ensembles for those cases where three states are required in
Sec.\ \ref{sec:three}. In Sec.\ \ref{sec:calc} we show how our results may be
applied to the calculation of the Holevo capacity. Conclusions are given in
Sec.\ \ref{sec:conc}.

\section{Two state ensembles}
\label{sec:2st}

To obtain the results, we use the representation of the qubit channel on the
Bloch sphere. A general qubit density operator may be expressed as
\begin{equation}
\rho = \frac 12 (\openone + \vec r \cdot \vec \sigma ),
\end{equation}
where $\vec\sigma$ is the vector of Pauli operators $(\sigma_x,\sigma_y,
\sigma_z)^{\rm T}$. The length of the vector $\vec r$ does not exceed 1, and its
components give the position of the state in the Bloch sphere. A qubit channel
$\Phi$ maps the sphere of possible input states to an ellipsoid, and may be
expressed as
\begin{equation}
\Phi(\rho) = \frac 12 [\openone + (\mathbf{\Lambda} \vec r + \vec t) \cdot
 \vec \sigma ].
\end{equation}
That is, the channel $\Phi$ produces the mapping $\vec r \mapsto
\mathbf{\Lambda} \vec r + \vec t$. Via local unitary operations
before and after the map, the transformation matrices $\mathbf{\Lambda}$ and
$\vec t$ may be brought to the form \cite{kingrus}
\begin{equation}
\label{em}
\mathbf{\Lambda} = \left( \begin{array}{*{20}c}
   \lambda_1 & 0 & 0 \\
   0 & \lambda_2 & 0 \\
   0 & 0 & \lambda_3 \\
\end{array}
\right), \qquad
\vec t = \left( \begin{array}{*{20}c}
   t_1 \\
   t_2 \\
   t_3 \\
\end{array}
\right).
\end{equation}

That is, an arbitrary qubit channel $\Phi$ may be expressed as $\Phi = \Gamma_U
\circ \Phi_{t,\Lambda} \circ \Gamma_V$, where $\Gamma_U$ and $\Gamma_V$ are
unitary channels, and $\Phi_{t,\Lambda}$ is the channel with $\mathbf{\Lambda}$
and $\vec t$ given by \eqref{em}. For this study, we consider the restricted
case of channels $\Phi$ such that the $x$ and $y$ components of $\vec t$ are
zero, and use the notation $t=t_3$. Hence $\vec t$ is given by
\begin{equation}
\label{dee}
\vec t = \left( \begin{array}{*{20}c}
   0 \\
   0 \\
   t \\
\end{array}
\right).
\end{equation}

In order to evaluate the Holevo capacity, we use an approach similar to that of
Ref.\ \cite{cortese}. The Holevo capacity may be given by the following
expression \cite{schu,ohya}:
\begin{equation}
\label{minmax}
C(\Phi) = \min_{\psi_0} \, \max_{\rho_0} D(\Phi(\rho_0) \| \Phi(\psi_0)),
\end{equation}
where $D$ is the relative entropy
\begin{equation}
D(\rho \| \psi) = {\rm Tr} (\rho \log \rho - \rho \log \psi ).
\end{equation}
Throughout this paper we use the convention that ``$\log$'' and ``$\exp$'' are
base 2, and logarithms base $e$ are given as ``$\ln$''. The relative entropy can
be evaluated using the following useful result from \cite{cortese}:
\begin{equation}
\label{relat}
D(\rho \| \psi) = \frac 12 \left[ f(r)-\log(1-q^2)
-r\cos(\theta) f'(q) \right],
\end{equation}
where
\begin{align}
f(x)&=(1+x)\log(1+x)+(1-x)\log(1-x), \\
f'(x)&=\log\left( \frac{1+x}{1-x}\right) .
\end{align}
The Bloch vectors for $\rho$ and $\psi$ are $\vec r$ and $\vec q$, respectively,
and we also define $r=|\vec r|$, $q=|\vec q|$, $\cos(\theta)=\vec r \cdot
\vec q/rq$.

To evaluate the Holevo capacity, we consider the action of the simplified
channel $\Phi_{t,\Lambda}$. This channel has the same capacity as $\Phi$,
because unitary operations do not affect the capacity. The set of possible
output states from the channel $\Phi_{t,\Lambda}$ forms an ellipsoid centred on
the $z$ axis. The ellipsoid has a radius of $|\lambda_1|$ in the $x$ direction,
and a radius of $|\lambda_2|$ in the $y$ direction. 

The nature of the optimal ensemble may be determined by considering the states
in the minmax formula \eqref{minmax}. In the following we take the states
$\rho=\Phi_{t,\Lambda}(\rho_0)$ and $\psi=\Phi_{t,\Lambda}(\psi_0)$ to be output
states from the simplified channel. If $\psi$ is the average output density
operator for an optimal ensemble, the operators $\rho_k$ that maximise
$D(\rho_k \| \psi)$ are possible output states for this ensemble. It is
necessary that there is some set of $p_k$ such that $\sum_k p_k \rho_k = \psi$.
The optimal ensemble is not necessarily unique, because there may be different
ways of choosing the probabilities such that $\sum_k p_k \rho_k = \psi$.
However, from Ref.\ \cite{cortese}, the optimal average output state is unique.

As we are restricting to operations such that $\vec t$ lies on the $z$ axis,
there are many simplifications due to the symmetry of the system. Many of these
simplifications were used in Ref.\ \cite{cortese} in the analysis of the
amplitude damping channel. We give a general explanation here. Firstly, the
optimal state $\psi$ must lie on the $z$ axis. To show this result, for any pair
of states $\rho$ and $\psi$, consider the second pair $\rho'$ and $\psi'$,
where $\vec r'=(-r_x,-r_y,r_z)^{\rm T}$ and $\vec q'=(-q_x,-q_y,q_z)^{\rm T}$.
Due to symmetry, if $\rho$ and $\psi$ are possible output states, then so are
$\rho'$ and $\psi'$. From the symmetry of the relative entropy, it is evident
that $D(\rho \| \psi)=D(\rho' \| \psi')$. This immediately implies that
$\max_\rho D(\rho \| \psi)=\max_\rho D(\rho \| \psi')$. Therefore, if $\psi$
minimises this quantity, then so does $\psi'$. However, as the optimal average
output state is unique, $\psi$ and $\psi'$ must coincide, which implies that
$\psi$ lies on the $z$ axis.

In the case that $|\lambda_1|\ne|\lambda_2|$, the $\rho_k$ that maximise the
relative entropy will lie in the $x-z$ plane if $|\lambda_1|>|\lambda_2|$, and
the $y-z$ plane if $|\lambda_1|<|\lambda_2|$. That is because $\psi$ lies on the
$z$ axis, so the relative entropy is symmetric under rotation about the $z$
axis. If $|\lambda_1|>|\lambda_2|$, then the ellipsoid has a radius in the $x$
direction larger than the radius in the $y$ direction. Consider any state $\rho$
that is not in the $x-z$ plane. We can determine a second state $\rho'$ in the
$x-z$ plane with Bloch vector $\vec r'=(\sqrt{r_x^2+r_y^2},0,r_z)^{\rm T}$. This
state is in the interior of the ellipsoid, and we may obtain a third state on
the surface of the ellipsoid, $\rho''$, by extending outwards in a straight line
from $\psi$. From Ref.\ \cite{cortese} (the first lemma in Sec.\ 5.3),
\begin{equation}
D(\rho''\|\psi) > D(\rho'\|\psi)=D(\rho\|\psi).
\end{equation}
This implies that $\rho$ does not maximise the relative entropy. Hence, all
$\rho_k$ that maximise the relative entropy must be in the $x-z$ plane.
Similarly, if $|\lambda_1|<|\lambda_2|$, the ellipsoid has a radius in the $y$
direction larger than the radius in the $z$ direction, and the optimal $\rho_k$
must be in the $y-z$ plane.

In the case that $|\lambda_1|=|\lambda_2|$, the situation is a little more
complicated. For each optimal $\rho_k$, there is a circle of optimal density
operators around the $z$ axis. However, in order to obtain an optimal ensemble,
it is only necessary to use non-zero probabilities such that $\sum_k p_k \rho_k
= \psi$. As $\psi$ lies on the $z$ axis, it is sufficient to take $\rho_k$ from
a single plane in the Bloch sphere that contains the $z$ axis.

This reasoning means that, regardless of the relative values of $|\lambda_1|$
and $|\lambda_2|$, we may restrict to considering $\rho_k$ that maximise
$D(\rho_k\|\psi)$ in a single plane in the Bloch sphere. Caratheodory's theorem
implies that there need be no more than three states in the ensemble. This fact
was also noted in Ref.\ \cite{kingnath}. The examples given by
Ref.\ \cite{hayashi} which needed four states used $\vec t$ that were not on the
$z$ axis.

In fact, in some cases the number of states required is only two \cite{fuchs},
though in some cases three are required. Here we give criteria that can show
when only two states are required via the following theorem:
\begin{thm} \label{th1} For a CPTP map $\Phi = \Gamma_U \circ
\Phi_{t,\Lambda} \circ \Gamma_V$ with $\mathbf{\Lambda}$ given by \eqref{em} and
$\vec t$ given by \eqref{dee}, if $\lambda_m=|\lambda_3|$ or $A\notin(0,1/2)$,
where
\begin{equation}
\label{ineq}
A=\frac {t^2\lambda_3^2}{\lambda_m^2-\lambda_3^2} -1+\lambda_m^2+t^2
\end{equation}
and $\lambda_m=\max(|\lambda_1|,|\lambda_2|)$, then there is an ensemble that
gives the maximum output Holevo information and has two states. 
\end{thm}

Before we proceed to the proof, we give some explanation of the quantity $A$.
Let us consider the output ellipse in the $x-z$ plane if $|\lambda_1|\ge
|\lambda_2|$, or the $y-z$ plane if $|\lambda_1|<|\lambda_2|$. A point on the
surface of this ellipse has a distance from the origin $r$, which is given by
Eq.\ \eqref{distance} in the proof below. Taking the derivative of $r^2$ with
respect to $\phi$ gives
\begin{equation}
\frac{d^2}{d\phi^2}(r^2) = 2\sin\phi \left[ (\lambda_m^2-\lambda_3^2)\cos\phi
 -\lambda_3 t \right].
\end{equation}
This expression is zero if $\sin\phi=0$, $\lambda_m^2-\lambda_3^2=\lambda_3t=0$,
or 
\begin{equation}
\cos\phi = \frac{\lambda_3t}{\lambda_m^2-\lambda_3^2}.
\end{equation}
The third case is only possible if the absolute value of the right-hand side
(RHS) does not exceed 1. If it does not, then substituting this expression for
$\cos\phi$ into the expression for $r$ gives the extremum
\begin{equation}
r_{\rm ex}^2 = \frac {t^2\lambda_3^2}{\lambda_m^2-\lambda_3^2}+\lambda_m^2+t^2
 = A+1.
\end{equation}
Therefore, in this case, $A$ is the difference between the square of an extremum
of $r$ and 1. In the case $\lambda_m^2-\lambda_3^2=\lambda_3t=0$, the radius is
independent of $\phi$. This possibility will be excluded in the discussion of
$A$, because $\lambda_m=|\lambda_3|$ is an alternative criterion to
$A\notin(0,1/2)$, and leads to infinite $A$.

If $A$ were positive, then $r_{\rm ex}^2$ would be larger than one, which is not
possible for CPTP maps. Therefore, for any map such that an extremum of $r$ is
obtained for $\sin\phi\ne 0$ (and $\lambda_m\ne|\lambda_3|$), the condition
$A\notin(0,1/2)$ is automatically satisfied due to the fact that states can not
mapped outside the Bloch sphere. However, $A\notin(0,1/2)$ is not satisfied for
every possible CPTP map, because for some $|\lambda_3t/(\lambda_m^2-
\lambda_3^2)|>1$.

Another case where $A\notin(0,1/2)$ is automatically satisfied is when
$\lambda_m<|\lambda_3|$. That is because the condition that the map is CPTP
implies that $\lambda_m^2+t^2\le 1$, and if $\lambda_m<|\lambda_3|$ then
$t^2\lambda_3^2/(\lambda_m^2-\lambda_3^2)$ is negative. Therefore, from the
definition of $A$, it is clear that $A\le 0$. We now proceed to the proof of the
theorem.

\begin{proof} We begin the analysis by mentioning some trivial cases that would
otherwise complicate the analysis. If $t=0$, then the channel is unital, and the
result in this case was proven in Ref.\ \cite{kingrus}. If all three of the
$\lambda_k$ are zero, then the channel capacity is zero, and the result is
trivial. If two of the $\lambda_k$ are zero, then the possible output states
form a line in the Bloch sphere, and the result follows from the fact that there
are only two extremal output states.

The result is also trivial if $\lambda_3=0$. In that case, since we may restrict
to considering states in the $x-z$ or $y-z$ plane, the set of output states that
it is sufficient to consider forms a line. The result again follows from the
fact that there are only two extremal states. For the remainder of the analysis
we take $t\ne 0$, $\lambda_3\ne 0$, and assume that no more than one of the
$\lambda_k$ is zero. This third assumption means that $\lambda_m\ne 0$.

For the remainder of this proof we consider the input and output states for the
simplified channel $\Phi_{t,\Lambda}$. The input and output states for the total
channel $\Phi$ will simply be rotated from these states. We take the input state
to have $\vec r=(\sin\phi,0,\cos\phi)^{\rm T}$ for $|\lambda_1|\ge|\lambda_2|$,
or $\vec r=(0,\sin\phi,\cos\phi)^{\rm T}$ for $|\lambda_1|<|\lambda_2|$. The
output state will then have $\vec r=(\lambda_1\sin\phi,0,t+
\lambda_3\cos\phi)^{\rm T}$ or $\vec r=(0,\lambda_2\sin\phi,t+
\lambda_3\cos\phi)^{\rm T}$. The state $\psi$ has $\vec q = (0,0,q_z)^{\rm T}$.
In either case, we have for the output
\begin{align}
\label{distance}
r &= \sqrt{\lambda_m^2\sin^2\phi+(t+\lambda_3\cos\phi)^2}, \nn \\
r\cos\theta &= (t+\lambda_3\cos\phi)\times{\rm sign}(q_z).
\end{align}
To search for the optimal $\rho$, it is merely necessary to search for the
optimal $\phi$. Because ${\rm sign}(q_z)f'(q)=f'(q_z)$, we may write the
relative entropy as
\begin{equation}
D(\rho\|\psi) = \frac 12 \left[ f(r) - \log(1-q_z) -
(t+\lambda_3\cos\phi)f'(q_z)\right].
\end{equation}

The derivative of $D(\rho\|\psi)$ with respect to $\phi$ is
\begin{align}
&\frac{d}{d\phi}D(\rho\|\psi) = \frac 12 \left\{ \frac{dr}{d\phi} f'(r) -
 f'(q_z) \frac{d}{d\phi} [t+\lambda_3\cos(\phi)] \right\} \nn \\
&= \frac 12 \left\{ [(\lambda_m^2-\lambda_3^2)\cos\phi -t\lambda_3]
 f'(r)/r + f'(q_z) \lambda_3 \right\}\sin\phi .
\end{align}
There will be extrema of $D(\rho\|\psi)$ for $\phi=0$ and
$\phi=\pi$, as well as when
\begin{equation}
\label{root}
[(\lambda_m^2-\lambda_3^2)\cos\phi -t\lambda_3] f'(r)/r = -f'(q_z) \lambda_3.
\end{equation}
We will consider the solutions of this equation for $\phi$ in the interval
$(0,\pi)$. Any solution in $(0,\pi)$ will yield a corresponding solution in
$(-\pi,0)$ due to symmetry.

Taking the derivative of the left-hand side (LHS) gives
\begin{align}
&\frac{d}{d\phi} [(\lambda_m^2-\lambda_3^2)\cos\phi -t\lambda_3] f'(r)/r =
 \left\{ - (\lambda_m^2-\lambda_3^2) \frac {f'(r)}r \right. \nn \\ &\left.
+[(\lambda_m^2-\lambda_3^2)\cos\phi - t\lambda_3]^2
\frac 1r \frac{d}{dr}\left( \frac {f'(r)}r \right)
\right\} \sin\phi.
\end{align}
In the case that $|\lambda_m|\ne |\lambda_3|$,
\begin{equation}
[(\lambda_m^2-\lambda_3^2)\cos\phi - t\lambda_3]^2 =
 (\lambda_m^2-\lambda_3^2)(1-r^2+A).
\end{equation}
We then obtain
\begin{align}
\label{nega}
& \frac{d}{d\phi} [(\lambda_m^2-\lambda_3^2)\cos\phi -t\lambda_3] f'(r)/r \nn \\
&=\frac{(\lambda_m^2-\lambda_3^2)\sin\phi}r [ h(r) +Ag(r)],
\end{align}
where
\begin{align}
g(r)&=\frac{d}{dr}\left( \frac {f'(r)}r \right) = \frac 2{(1-r^2)r\ln 2} -
\frac{1}{r^2}\log\left(\frac{1+r}{1-r}\right), \\
h(r)&=\frac 2r - \frac{f'(r)}{r^2} = \frac 2{r\ln 2}-
\frac{1}{r^2}\log\left(\frac{1+r}{1-r}\right).
\end{align}
The functions $g(r)$ and $h(r)$ satisfy the inequalities
\begin{equation}
g(r)>0, \quad h(r)<0, \quad 2h(r)+g(r)>0,
\end{equation}
for $r\in (0,1)$. If $A\le 0$, then $h(r)+Ag(r)$ is negative for $r\in (0,1)$.
Similarly, if $A\ge 1/2$, then $h(r)+Ag(r)$ is positive for $r\in (0,1)$. In
either case $h(r)+Ag(r)$ has constant sign. We do not need to consider the
possibility that $r=0$, because this value is only possible when $\sin\phi=0$
(for $\lambda_m\ne 0$).

The case where $r=1$ is more complicated. It is possible for $r$ to be equal to
1 for $\phi\in(0,\pi)$. In the case where $r$ has a maximum for
$\phi\in(0,\pi)$, the maximum value of $r$ is $A+1$. If $r$ is equal to 1 for
$\phi\in(0,\pi)$, this must be a maximum, and therefore $A=0$ (as we are taking
$\lambda_m\ne|\lambda_3|$). That implies that the expression in square brackets
on the LHS of Eq.\ \eqref{root} is proportional to $\sqrt{1-r^2}$. Hence
the LHS of \eqref{root} approaches zero as $r$ approaches 1, and is continuous
as a function of $\phi$ for $\phi\in(0,\pi)$. As $h(r)+Ag(r)$ has constant sign
for all values of $\phi\in(0,\pi)$ except where $r=1$, and the LHS of
\eqref{root} is continuous where $r=1$, the LHS of \eqref{root} is one-to-one in
this interval.

For the case $\lambda_m=|\lambda_3|$,
\begin{equation}
\label{equal}
\frac{d}{d\phi} [(\lambda_m^2-\lambda_3^2)\cos\phi -t\lambda_3] f'(r)/r =
t^2\lambda_3^2 (\sin\phi) g(r)/r .
\end{equation}
Therefore, the derivative of the LHS of \eqref{root} is nonzero for
$\phi\in(0,\pi)$. Note that we are assuming that $t\ne 0$ and $\lambda_3\ne 0$,
so the RHS of Eq.\ \eqref{equal} is nonzero. Thus we have shown that, regardless
of the relative values of $\lambda_m$ and $\lambda_3$, the LHS of \eqref{root}
is a one-to-one function of $\phi$, and there can be at most one solution of
\eqref{root} in $(0,\pi)$. If there is a solution, it must correspond to an
extremum, because a point of inflection would conflict with the fact that the
LHS of \eqref{root} is one-to-one.

As $D(\rho\|\psi)$ is symmetric about $\phi=0$, there must be two solutions of
\eqref{root} with $\sin\phi\ne 0$ or none. In the case where there are no
solutions, there are only two extrema (for $\phi=0$ and $\pi$), and only one of
these can be a maximum. This is not consistent with $\psi$ being optimal,
because the optimal ensemble can not have only one state. Therefore, if $\psi$
is optimal, then there must be two solutions of \eqref{root}. As the maxima and
minima alternate, the maxima are either at $\phi=0$ and $\pi$, or the solutions
of \eqref{root}.

In the case that $|\lambda_1|\ne|\lambda_2|$, this result immediately implies
that there are only two states in the optimal ensemble. In the case
$|\lambda_1|=|\lambda_2|$, if the maxima correspond to the solutions of
\eqref{root}, optimal ensembles may contain any states in a ring about the $z$
axis. However, as discussed above, it is only necessary to consider $\rho_k$ in
one plane in the Bloch sphere in this case, so there is again an optimal
ensemble with two members. \end{proof}

It is also possible to determine simple criteria for when the optimal states in
the ensemble are on the $z$ axis, and when the optimal states in the ensemble
correspond to the maxima for $\sin\phi\ne 0$. The result is:
\begin{thm} \label{th2} Let $\Phi_{t,\Lambda}$ be a CPTP map with
$\mathbf{\Lambda}\ne 0$ given by \eqref{em} and $\vec t$ given by \eqref{dee}.
The condition that $\lambda_m=|\lambda_3|$ or $A\notin(0,1/2)$ may be expressed
as two alternative mutually exclusive conditions: \\
{\rm Condition 1}. $\lambda_m \le |\lambda_3|$ or $A\ge 1/2$ \\
{\rm Condition 2}. $\lambda_m>|\lambda_3|$ and $A\le 0$ \\
If Condition 1 is satisfied, the optimal ensemble consists of two states on the
$z$ axis. If Condition 2 is satisfied, there is an optimal ensemble consisting
of two states equidistant from the $z$ axis and lying on a line perpendicular
to and intersecting the $z$ axis. \end{thm}

Here we have given the result in terms of the simplified map $\Phi_{t,\Lambda}$,
rather than expressing it in terms of the arbitrary map $\Phi$. That is because
the ellipse of output states will be rotated for the arbitrary map, so it is not
possible to express the result in this way. The statement of this theorem also
differs in that $\mathbf{\Lambda}$ is taken to be non-zero. This is to exclude
the trivial case where all ensembles give zero Holevo information.

\begin{proof} As was shown above, $\lambda_m<|\lambda_3|$ also implies that
$A\le 0$. Another consequence of this is that, if $A>0$, then $\lambda_m>
|\lambda_3|$. Therefore Condition 1 contains three alternatives: \\
1. $\lambda_m = |\lambda_3|$ \\
2. $\lambda_m < |\lambda_3|$ and $A\le 0$ \\
3. $A\ge 1/2$ and $\lambda_m > |\lambda_3|$ \\
It is clear that, for each of these three alternatives, the conditions of
Theorem \ref{th1} must hold. If none of these alternatives apply, but
$A\notin(0,1/2)$, then $\lambda_m>|\lambda_3|$ and $A\le 0$, which is Condition
2 given in the theorem.

To determine which extrema of $D(\rho\|\psi)$ are maxima and which are minima,
it is sufficient to consider the point $\phi=0$. At this point, the second
derivative of $D(\rho\|\psi)$ is given by
\begin{equation}
\frac{d^2}{d\phi^2}D(\rho\|\psi) =\frac 12 \big\{ [(\lambda_m^2-
\lambda_3^2)-t\lambda_3]f'(r)/r+f'(q_z) \lambda_3 \big\}.
\end{equation}
We know that the LHS of \eqref{root} is one-to-one, and there must be at least
one solution of \eqref{root} if $\psi$ is optimal (otherwise there would be only
one possible state for the ensemble).

If $\lambda_m = |\lambda_3|$, then from \eqref{equal}, the LHS of \eqref{root}
is monotonically increasing for $\phi\in(0,\pi)$. If $A\ge 1/2$ and
$\lambda_m > |\lambda_3|$, then $h(r)+Ag(r)>0$, and from \eqref{nega} the LHS of
\eqref{root} is monotonically increasing. Similarly, if
$\lambda_m < |\lambda_3|$ and $A\le 0$, then $h(r)+Ag(r)<0$, and the LHS of
\eqref{root} is again monotonically increasing. Therefore, for all three
alternatives for Condition 1, the LHS of \eqref{root} is monotonically
increasing for $\phi\in(0,\pi)$. For Condition 2, $\lambda_m > |\lambda_3|$ and
$A\le 0$, so $h(r)+Ag(r)<0$, and the LHS of \eqref{root} is monotonically
decreasing for $\phi\in(0,\pi)$.

If the LHS of \eqref{root} is monotonically increasing for $\phi\in(0,\pi)$, the
LHS of \eqref{root} must be less than the RHS for $\phi=0$, so
\begin{equation}
[(\lambda_m^2-\lambda_3^2)-t\lambda_3] f'(r)/r + f'(q_z) \lambda_3 < 0.
\end{equation}
This means that the second derivative of $D(\rho\|\psi)$ is negative for
$\phi=0$, and $D(\rho\|\psi)$ is a maximum at this point. Hence, the two maxima
are obtained for $\phi=0$ and $\pi$, and these values correspond to the states
in the optimal ensemble. Thus we see that, for Condition 1, the LHS of
\eqref{root} is monotonically increasing and the optimal ensemble consists of
two states on the $z$ axis.

Alternatively, for Condition 2, the LHS of \eqref{root} is monotonically
decreasing, so the LHS of \eqref{root} is greater than the RHS for $\phi=0$,
and less for $\phi=\pi$. This implies that the second derivative of
$D(\rho\|\psi)$ is positive for $\phi=0$ and $\phi=\pi$, and these points are
minima. Hence, in this case the states in the optimal ensemble correspond to
the extrema of $D(\rho\|\psi)$ for $\sin\phi\ne 0$. 

In the case that $|\lambda_1|>|\lambda_2|$ or $|\lambda_1|<|\lambda_2|$, the
optimal ensemble must be in the $x-z$ plane or $y-z$ plane, respectively. In
either case, two maxima are obtained in the appropriate plane for $\phi=\pm
\phi_0$, where $\phi_0$ maximises $D(\rho\|\psi)$. These two solutions are
equidistant from the $z$ axis, and on a line perpendicular to and intersecting
the $z$ axis. If $|\lambda_1|=|\lambda_2|$, then there will be a circle of
states about the $z$ axis that maximise the relative entropy. Optimal ensembles
may contain any number of these states. However, as discussed above we may
restrict to states in one plane. This yields an ensemble with two members that
again lie on a line perpendicular to and intersecting the $z$ axis. \end{proof}

Another issue is the position of the optimal average output state. It is
possible to use similar techniques as above to show that this state should be
further from the centre of the Bloch sphere than the output for the maximally
mixed state. Specifically, $q_z$ for the optimal average output state should
satisfy $q_z/t>1$ for $t$ and $\lambda_3$ both nonzero. The case $t=0$ means
that the map is unital, and it is known in that case that $q_z=0$ is optimal.
If $\lambda_3=0$, then clearly $q_z=t$.

To show this result, let us assume some value for $q_z$, (the other components
of $\vec q$ are zero), and take a value of $\phi$ such that
$|t+\lambda_3\cos\phi|>|t-\lambda_3\cos\phi|$. We denote the states with
$r_z=t\pm\lambda_3\cos\phi$ by $\rho_\pm$. Determining the difference in
relative entropies gives
\begin{align}
\label{difrel}
& D(\rho_+\|\psi)-D(\rho_-\|\psi) \nn \\
& \quad = f(r_+)-f(r_-)-2\lambda_3\cos\phi f'(q_z) \nn \\
& \quad > f'(\bar r)(r_+-r_-)-2\lambda_3\cos\phi f'(q_z),
\end{align}
where $r_\pm$ is the magnitude of the Bloch vector for $\rho_\pm$, and
$\bar r=(r_++r_-)/2$. In the second line we have used the strict convexity of
$f'(r)$ and the Hermite-Hadamard inequality \cite{hadamard}. Now using the fact
that $r_+^2-r_-^2=4t\lambda_3\cos\phi$, we have
$r_+-r_-=(2t\lambda_3\cos\phi)/\bar r$. Therefore Eq.\ \eqref{difrel}
simplifies to
\begin{equation}
D(\rho_+\|\psi)-D(\rho_-\|\psi) >
2t\lambda_3\cos\phi [f'(\bar r)/\bar r- f'(q_z)/t].
\end{equation}
We have chosen $\phi$ such that $t\lambda_3\cos\phi$ is positive, and both
$f'(x)$ and $f'(x)/x$ are monotonically increasing functions. Also $\bar r\ge
t$, with equality only if $\lambda_m\sin\phi=0$. Therefore, $q_z/t\le 1$
implies that
\begin{equation}
D(\rho_+\|\psi)-D(\rho_-\|\psi) > 0.
\end{equation}

This means that, if $t$ is positive and $q_z\le t$, then all states $\rho_-$
that have $z$ component of their Bloch vector less than $t$ do not maximise the
relative entropy. In addition, if $q_z=t$ the relative entropy can not be
maximised for $r_z=t$. In the case $\lambda_m=0$ this is trivial, because the
maxima are for $r_z=t+\lambda_3$ and $r_z=t-\lambda_3$. If $\lambda_m\ne 0$,
then $f'(r)/r>f'(t)/t$. As we are also taking $\lambda_3\ne 0$, this
inequality means that Eq.\ \eqref{root} can not be satisfied for $\phi=\pi/2$.

Hence, for $q_z\le t>0$ and $\lambda_3\ne 0$, all $\rho_k$ that maximise the
relative entropy must have a $z$ component of their Bloch vector greater than
that for $\psi$, and they can not give an average equal to $\psi$. This is not
consistent with $\psi$ being the average state for the optimal ensemble, and
therefore the average state for the optimal ensemble must satisfy $q_z>t$.
Similarly, if $t$ is negative and $\lambda_3\ne 0$, then the average state for
the optimal ensemble satisfies $q_z<t$.

With the aid of this result, we can alternatively express Theorem \ref{th2} in
terms of the orthogonality of the input states. The result is:
\begin{cor} Consider a CPTP map $\Phi = \Gamma_U \circ \Phi_{t,\Lambda}
\circ \Gamma_V$ with $\mathbf{\Lambda}\ne 0$ given by \eqref{em} and $\vec t$
given by \eqref{dee}. The condition that $\lambda_m=|\lambda_3|$ or
$A\notin(0,1/2)$ may be expressed as two alternative mutually exclusive
conditions: \\
{\rm Condition 1}. $\lambda_m \le |\lambda_3|$ or $A\ge 1/2$ \\
{\rm Condition 2}. $\lambda_m>|\lambda_3|$ and $A\le 0$ \\
If $t\ne 0$ and $\lambda_3\ne 0$, the maximum output Holevo information is
obtained for two orthogonal input states if Condition 1 is satisfied, and two
non-orthogonal input states if Condition 2 is satisfied. \end{cor}

\begin{proof} Note first that unitary operations do not change the orthogonality
relations between the states. Therefore it is sufficient to prove the
orthogonality relations for the simplified map $\Phi_{t,\Lambda}$. For
Condition 1 the result follows immediately from Theorem \ref{th2}. The two input
states are the extremal states on the $z$ axis, and therefore are $\ket 0$ and
$\ket 1$, which are orthogonal.

To prove the result for Condition 2, we use the result that, for $t\ne 0$ and
$\lambda_3\ne 0$, $q_z$ is not equal to $t$. If the input states for Condition 2
were orthogonal, then that would lead to $q_z=t$. Therefore, if $t\ne 0$ and
$\lambda_3\ne 0$, the input states must be non-orthogonal if Condition 2 holds.
\end{proof}

\section{Applications}
\label{sec:apply}
These results allow us to make sense of the results obtained in previous work.
In particular, \cite{cortese} found that only two states in the ensemble were
required for the amplitude damping channel, where $\lambda_1=\lambda_2=
\sqrt{\mu}$, $\lambda_3=\mu$ and $t=1-\mu$. We find that, in this case, $A=0$,
so $A\notin(0,1/2)$ is satisfied and Theorem \ref{th1} predicts that the optimal
ensemble requires two states. For this channel, $\lambda_m>|\lambda_3|$ and
$A\le 0$, which corresponds to Condition 2 in Theorem \ref{th2}. Theorem
\ref{th2} therefore predicts that, for this channel, the optimal ensemble
consists of two states at the same distance from the $x-y$ plane, rather than on
the $z$ axis. This is what was found in Ref.\ \cite{cortese}.

Another channel is the shifted depolarising channel, which was considered in
Ref.\ \cite{kingnath}. For this channel, $\lambda_k=\mu$ and $t=1-\mu$. As
$\lambda_m=\lambda_3$, Theorem \ref{th1} applies, and the ensemble should require only
two states. This result is what was found in \cite{kingnath}. Also, because
$\lambda_m=\lambda_3$, Condition 1 in Theorem \ref{th2} holds, so Theorem
\ref{th2} predicts that the states in the optimal ensemble lie on the $z$ axis.
This is also consistent with the results of Ref.\ \cite{kingnath}.

On the other hand, let us consider the examples given in \cite{kingnath} that
require three states. For one of these examples, $\lambda_1=\lambda_2=0.6$ and
$\lambda_3=t=0.5$, so $A\approx 0.178$. This is in the interval $(0,1/2)$, so it
is not surprising that three states are required. Another example is
$\lambda_1=t=0.5$ and $\lambda_2=\lambda_3=0.435$; in this case $A$ is about
0.278, which is again in the interval $(0,1/2)$.

In Ref.\ \cite{kingnath} a strategy used to find channels that require three
states was to vary the parameters from a channel such that the optimal states
are on the $z$ axis to one where the optimal states are away from the $z$ axis.
This strategy can alternatively be explained in terms of Theorem \ref{th2}. The
channel parameters can not be continuously varied from Condition 1 to Condition
2 without $A$ passing through the interval $(0,1/2)$. That is because it is not
possible to continuously vary the channel parameters from
$\lambda_m<|\lambda_3|$ to $\lambda_m>|\lambda_3|$ while maintaining the same
sign for $A$.

\begin{figure}
\centering
\includegraphics[width=0.45\textwidth]{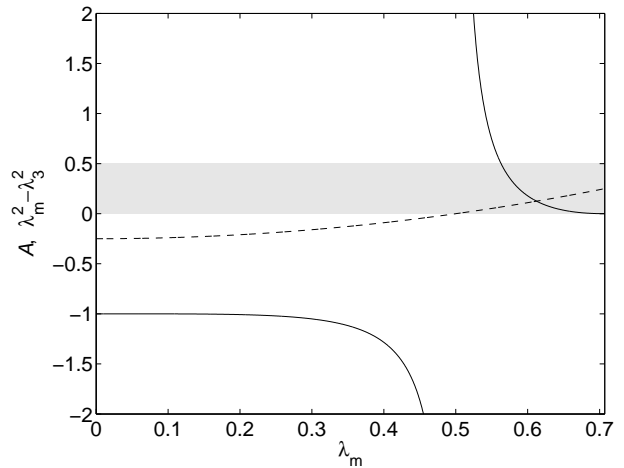}
\caption{The values of $A$ (solid line) and $\lambda_m^2-\lambda_3^2$ (dashed
line) as a function of $\lambda_m$ for $\lambda_3=t=1/2$. The shaded region
shows the region of values of $A$ such that the optimal ensemble may require
three states. Results for $\lambda_m>1/\sqrt{2}$ are not shown, because the maps
for $\lambda_m>1/\sqrt{2}$ are not CPTP.}
\label{fig1}
\end{figure}

To take an example from \cite{kingnath}, let $\lambda_3=t=1/2$, and vary
$\lambda_m$. Then the variation of $A$ and $\lambda_m^2-\lambda_3^2$ are as in
Fig.\ \ref{fig1}. It can be seen from this figure that as $\lambda_m^2-
\lambda_3^2$ passes through zero, $A$ switches from negative to positive. In
fact the only point where Condition 2 is satisfied is for $\lambda_m=1/\sqrt
2$. In passing from $\lambda_m=0.5$, where $\lambda_m=\lambda_3$, to
$\lambda_m=1/\sqrt 2$, the value of $A$ passes through $(0,1/2)$.

A case of particular interest is that where $\mathbf{\Lambda}$ and $\vec t$ are
given by
\begin{equation}
\label{extremal}
\mathbf{\Lambda} = \left( \begin{array}{*{20}c}
   \cos \delta & 0 & 0 \\
   0 & \cos \gamma & 0 \\
   0 & 0 & \cos \gamma \cos \delta \\
\end{array}
\right), \quad
\vec t = \left( \begin{array}{*{20}c}
   0 \\
   0 \\
   \sin \gamma \sin \delta \\
\end{array}
\right).
\end{equation}
This type of channel arises naturally when considering qubit interactions. If
one introduces an ancilla qubit, performs a unitary operation, then traces over
this ancilla qubit, the resulting operation is of this form \cite{niu}. Maps of
this form also arise naturally when considering extremal maps \cite{ruskai}.
Also, it is known that all qubit maps with two Kraus operators are of this form
\cite{ruskai}.

For maps of this form, we find that $A=0$, so the conditions of Theorem
\ref{th1} are satisfied. Therefore, for maps that arise from a unitary
interaction with an ancilla qubit, the optimal ensemble requires only two
states. This result was also claimed in Ref.\ \cite{verst}, although the
complete proof was not given. In addition, $|\lambda_3|< \lambda_m$, so from
Theorem \ref{th2} the two states for the optimal ensemble are away from the $z$
axis.

\section{Three state ensembles}
\label{sec:three}

In the case where three states are required for the optimal ensemble, it is
possible to show that one of the states needs to be on the $z$ axis. The result
is
\begin{thm} \label{th3} Consider a CPTP map $\Phi_{t,\Lambda}$
with $\mathbf{\Lambda}$ given by \eqref{em} and $\vec t$ given by \eqref{dee}.
If the Holevo capacity can not be achieved with a two-state ensemble, then any
optimal ensemble with three states consists of one state on the $z$ axis, and
two states equidistant from the $z$ axis and on a line perpendicular to and
intersecting the $z$ axis. The optimal input state on the $z$ axis is $\ket 0$
if $|t+\lambda_3|>|t-\lambda_3|$, and $\ket 1$ if $|t+\lambda_3|<|t-\lambda_3|$.
\end{thm}

\begin{proof} In order to prove the result, we start by considering the
expression in square brackets in \eqref{nega}. Although $h(r)+Ag(r)$ can change
sign, it is only zero for one value of $r$. To show this result, we use the
following facts:
\begin{align}
h(r) < 0, \quad g(r) > 0, \quad & g'(r) > 0, \\
h'(r)g(r)-g'(r)h(r) &> 0 .
\end{align}
These inequalities are all for $r\in(0,1)$, and are easily checked by plotting
the functions. If $h(r)+Ag(r)\ge 0$ for $r=r_0$, then $A\ge -h(r_0)/g(r_0)$, so
$h'(r_0)+Ag'(r_0)\ge(h'(r_0)g(r_0)-g'(r_0)h(r_0))/g(r_0)>0$. Therefore, if
$h(r)+Ag(r)\ge 0$ for $r=r_0$, then $h(r)+Ag(r)$ is increasing for $r=r_0$. This
implies that, if there is a value of $r$ for which $h(r)+Ag(r)=0$, then
$h(r)+Ag(r)>0$ for all larger values of $r$. Hence $h(r)+Ag(r)$ can be zero for
only one value of $r$ in $(0,1)$.

Recall that, if there is an extremum of $r$ for $\sin\phi\ne 0$, then the
condition $A\ne(0,1/2)$ is satisfied, and therefore the optimal ensemble
requires no more than two states. In the conditions for Theorem \ref{th3}, the
optimal ensemble requires more than two states, so $r$ has no extremum for
$\sin\phi\ne 0$. Hence $r$ is a one-to-one function for $\phi$ in the interval
$(0,\pi)$. Combining this result with the above reasoning, the RHS of
\eqref{nega} can be zero for only one value of $\phi$ in the interval $(0,\pi)$.

These results imply that the LHS of \eqref{root} can have a turning point for
only one value of $\phi$ in $(0,\pi)$, and therefore there are at most two
solutions of \eqref{root} for $\phi\in(0,\pi)$. In turn this implies that there
are no more than two extrema of $D(\rho\|\psi)$ for $\phi\in(0,\pi)$. In fact,
there must be exactly two (if $\psi$ is optimal), because if there were only
one, then the optimal ensemble would require only two states, which violates the
conditions of Theorem \ref{th3}. 

Thus there will be two extrema of $D(\rho\|\psi)$ for $\phi\in(0,\pi)$, two
symmetric extrema for $\phi\in(-\pi,0)$, and extrema at $\phi=0$ and $\pi$.
These extrema must alternate between minima and maxima, and so one of the
extrema at $\phi=0$ and $\pi$ will be a maximum, and the other will be a
minimum. To determine which points are minima and which are maxima, consider the
second derivative of $D(\rho\|\psi)$ at a solution of \eqref{root}:
\begin{equation}
\frac{d^2}{d\phi^2}D(\rho\|\psi) = \frac{(\lambda_m^2-\lambda_3^2)\sin^2\phi}
{2r}[ h(r) +Ag(r)].
\end{equation}
Recall that, if $h(r)+Ag(r)$ is positive for $r=r_0$, it must also be positive
for $r>r_0$. Therefore, for the solution of \eqref{root} with smaller $r$,
$h(r)+Ag(r)$ is negative, and for the solution with larger $r$, $h(r)+Ag(r)$ is
positive.

For maps that require three states to achieve the Holevo capacity, $A>0$. As
discussed above, this implies that $\lambda_m>|\lambda_3|$, so $\lambda_m^2-
\lambda_3^2$ is positive. Thus multiplication by $\lambda_m^2-\lambda_3^2$ does
not change the sign, so the solution of \eqref{root} with smaller $r$ is a
maximum, and the solution with larger $r$ is a minimum. As the extrema alternate
between maxima and minima, the extremum on the $z$ axis that is closer to the
origin must be a minimum. Therefore, if $|t+\lambda_3|$ is greater than
$|t-\lambda_3|$, then the optimal output state on the $z$ axis will be at
$t+\lambda_3$. This corresponds to an input state of $\ket 0$. Similarly, if
$|t-\lambda_3|$ is greater than $|t+\lambda_3|$, then the optimal output state
on the $z$ axis is at $t-\lambda_3$, which corresponds to the input state
$\ket 1$.

The two remaining states in the optimal ensemble will correspond to solutions
$\phi=\pm\phi_0$ of \eqref{root}. In the case that $|\lambda_1|\ne|\lambda_2|$,
these states are in the $x-z$ or $y-z$ plane of the Bloch sphere, depending on
whether $|\lambda_1|>|\lambda_2|$ or $|\lambda_1|<|\lambda_2|$. In either case
the states are equidistant from the $z$ axis, on a line that is perpendicular to
and intersecting the $z$ axis. If $|\lambda_1|=|\lambda_2|$, then optimal
ensembles may contain any states from a circle about the $z$ axis. However, for
optimal ensembles with three states, the condition that the mean state is on the
$z$ axis restricts the remaining two states to be equidistant from the $z$ axis,
and on a line perpendicular to and intersecting the $z$ axis. \end{proof}

\section{Calculating capacities}
\label{sec:calc}
These results enable us to determine numerically efficient ways of calculating
capacities. In the case that the channel satisfies the conditions of Theorem
\ref{th1}, the problem becomes particularly simple. First it is necessary to
check whether it is Condition 1 or Condition 2 in Theorem \ref{th2} that is
satisfied. For Condition 1, the optimal ensemble consists of the two extremal
states on the $z$ axis. The probabilities may be determined by the fact that
$D(\rho_1\|\psi)=D(\rho_2\|\psi)$. The expression for the relative entropy
\eqref{relat} simplifies to
\begin{equation}
\label{analytic}
D(\rho\|\psi) = \frac 12 \left[ f(r_z) - \log (1-q_z^2) - r_z f'(q_z) \right].
\end{equation}
The condition that $D(\rho_1\|\psi)=D(\rho_2\|\psi)$ then becomes
\begin{align}
f(t+\lambda_3) - (t+\lambda_3) f'(q_z) = f(t-\lambda_3) - (t-\lambda_3)f'(q_z).
\end{align}
This may be solved for $q_z$, yielding
\begin{equation}
\label{answer}
q_z = \frac{X-1}{X+1},
\end{equation}
where
\begin{equation}
X = \exp \left[ \frac{f(t+\lambda_3)-f(t-\lambda_3)}{2\lambda_3} \right].
\end{equation}
Recall that we are using notation where ``exp'' means 2 to the power of the
argument. The channel capacity is obtained by substituting \eqref{answer} into
\eqref{analytic}. Thus the channel capacity may be obtained analytically. The
optimal ensemble may also be determined analytically. The optimal states
correspond to points on the $z$ axis at $t\pm\lambda_3$, and the probabilities
are given by
\begin{equation}
p_{\pm} = \frac 12 \pm \frac{q_z-t}{2\lambda_3}.
\end{equation}

For Condition 2 in Theorem \ref{th2}, the optimal states are away from the $z$
axis. Because $\psi$ must be the average of the two $\rho_k$, and the $z$
components of the two $\vec r_k$ are equal, the $z$ component of $\vec q$ must
also be equal. If $\psi$ is optimal, for the solution of \eqref{root} the $z$
component of $r$ should be equal to the $z$ component of $q$. Therefore the
optimal ensemble may be found by finding the solution of \eqref{root} with
$q_z=r_z$. Thus finding the capacity in this case reduces to finding the zero of
a function of a single real variable, which is easily performed numerically.

As an alternative interpretation of this result, consider the ensemble
consisting of two states corresponding to $\phi=\pm\phi_0$. The Holevo
information of this ensemble is given by
\begin{equation}
\label{maxim}
D(\rho_{\pm}\|\psi) = \frac 12 \left[ f(r)-f(r_z)\right],
\end{equation}
where $\psi$ is the average state. If the optimal ensemble is of this form, then
the maximum of this quantity gives the Holevo capacity for the channel. Taking
the derivative with respect to $\phi$, we find that the maximum will be for a
solution of \eqref{root} with $q_z=r_z$.

For the case where $\mathbf{\Lambda}$ and $\vec t$ are as given in
\eqref{extremal}, the problem of calculating the capacity has been considered in
Ref.\ \cite{uhlmann}. For this case, this reference gives an analytic method for
calculating the Holevo capacity for given mean state. Although this method was
derived in quite a different way than the method given here, it is equivalent.

In those cases where $A\in(0,1/2)$, it is still possible that two states may be
sufficient for the optimal ensemble. In those cases, the ensemble must still
consist of either two states on the $z$ axis of the Bloch sphere, or two states
corresponding to $\phi=\pm\phi_0$, where $\phi_0$ is a root of \eqref{root}.
This result may be shown by considering $D(\rho\|\psi)$ as a function of $\phi$.
As was shown in the previous section, there can be at most three maxima of
$D(\rho\|\psi)$. If there are only two, then these are at $\phi=0$ and $\pi$ or
$\phi=\pm\phi_0$. In either case, the form of the optimal ensemble is the same
as for channels satisfying the conditions of Theorem \ref{th1}.

If there are three maxima, then one of these is on the $z$ axis, and the other
two are for $\phi=\pm\phi_0$. If two states are sufficient for the optimal
ensemble, these states must correspond to $\phi=\pm\phi_0$, because otherwise
$\psi$ would not be on the $z$ axis. Therefore, regardless of whether there are
two maxima or three, if two states are sufficient for the optimal ensemble, then
these consist of either two states on the $z$ axis, or two states corresponding
to $\phi=\pm\phi_0$.

These results can be used to determine if the optimal ensemble requires three
states in cases where $A\in(0,1/2)$. From the ``sufficiency of maximal distance
property'' in \cite{schu}, we know that the ensemble is optimal if there are no
values of $\rho$ that give values of $D(\rho\|\psi)$ greater than the $\rho_k$
in the ensemble. Therefore, in order to determine if the ensemble requires more
than two states, determine $\psi$ via the two different methods above. If, for
one of them, $D(\rho\|\psi)$ is maximised for the corresponding $\rho_k$, then
the optimal ensemble requires only two states. If neither of these methods gives
the optimal ensemble, then we have eliminated all possibilities for optimal
two-state ensembles, and the optimal ensemble must require three states.

It is also possible to efficiently determine the Holevo capacity in those cases
where the ensemble requires three states. The reason for this is that the only
unknowns for the three state ensemble are the value of $\phi_0$ such that
$\phi=\pm\phi_0$ for the two off-axis states, and the probabilities for the
three states. Given the value of $\phi_0$, there is an analytic method to
determine the probabilities. Therefore the problem reduces to a numerical
maximisation in a single real variable, which is easily performed.

From Theorem \ref{th3}, the state on the $z$ axis will be at $t+\lambda_3$ if
$|t+\lambda_3|>|t-\lambda_3|$, and $t-\lambda_3$ if
$|t+\lambda_3|<|t-\lambda_3|$. Taking the other two states to correspond to
$\phi=\pm\phi_0$, the condition that the relative entropy $D(\rho_k\|\psi)$ is
independent of $k$ becomes
\begin{equation}
f(t\pm\lambda_3)-(t\pm\lambda_3) f'(q_z)= f(r_0)-(t+\lambda_3\cos\phi_0)f'(q_z),
\end{equation}
where $r_0^2=\lambda_1^2\sin^2\phi_0+(t+\lambda_3\cos\phi_0)^2$. We take the
plus sign if $|t+\lambda_3|>|t-\lambda_3|$, and the minus sign if
$|t+\lambda_3|<|t-\lambda_3|$. Solving for $q_z$ gives
\begin{equation}
q_z = \frac{X-1}{X+1},
\end{equation}
where
\begin{equation}
X = \exp \left[ \frac{f(t\pm\lambda_3)-f(r_0)}{\lambda_3(\pm 1-\cos\phi_0)}
\right].
\end{equation}
Note that this solution is reasonable only if the value of $q_z$ obtained is
between $t\pm\lambda_3$ and $t+\lambda_3\cos\phi_0$; otherwise negative
probabilities would be required for the ensemble.

Given this solution for $q_z$, the common value of the relative entropy is given
by
\begin{equation}
D(\rho_k\|\psi) = \frac 12 \left[ f(t\pm\lambda_3) - \log (1-q_z^2) -
 (t\pm\lambda_3) \log X \right].
\end{equation}
By finding the maximum of this (with $q_z$ between $t\pm\lambda_3$ and
$t+\lambda_3\cos\phi_0$), the Holevo capacity may be determined.

\begin{figure}[t]
\centering
\includegraphics[width=0.45\textwidth]{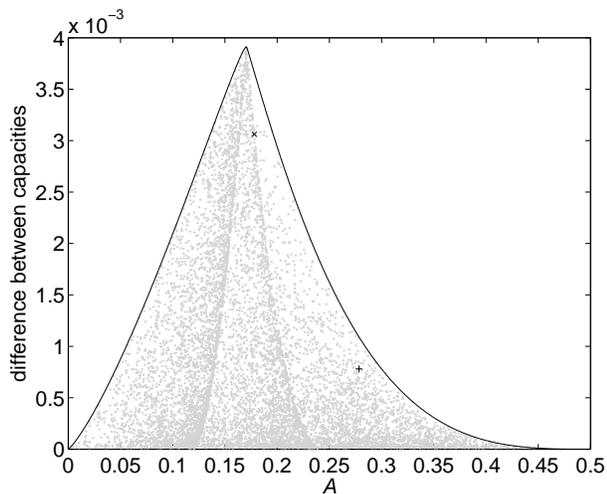}
\caption{The difference between the two-state capacity and the three-state
capacity versus the value of $A$. Random samples are shown as grey points, and
the numerically obtained upper bound is shown as the solid line. The cross and
plus are examples from Ref.\ \cite{kingnath}. The cross is for $\lambda_1=
\lambda_2=0.6$ and $\lambda_3=t=0.5$, and the plus is for $\lambda_1=t=0.5$ and
$\lambda_2=\lambda_3=0.435$.}
\label{fig2}
\end{figure}

This method was used to determine the difference between the two-state capacity
and the three-state capacity for a range of different maps. This difference is
plotted as a function of $A$ in Fig.\ \ref{fig2}. In addition, the states that
maximise this difference were searched for numerically for given values of $A$;
these results are also shown in Fig.\ \ref{fig2}. It can be seen that the
maximum difference in the capacities is still quite small; less than 0.004.
Also, the difference can be nonzero in the entire interval $(0,1/2)$. The
difference approaches zero quite rapidly as $A$ approaches $1/2$, but is still
nonzero. For comparison, two of the examples from Ref.\ \cite{kingnath} are
shown in Fig.\ \ref{fig2}. It was also found that, regardless of the value of
$A$, there were cases where two states were sufficient for the optimal ensemble.

\section{Conclusions}
\label{sec:conc}
We have shown a number of results on the form of optimal ensembles for qubit
channels. The class of channels considered includes those that can be
simplified, via unitary operations before and after the channel, to a form that
is symmetric under reflections in the $x-z$ and $y-z$ planes. This class
includes extremal channels, and most examples of channels considered in
previously published work. For these channels we have introduced the parameter
$A$, which can be interpreted in some cases in terms of the distance between
the output ellipsoid and the unit sphere.

The main result is that if $A$ is not in the interval $(0,1/2)$, then two states
are sufficient for the ensemble that maximises the Holevo capacity. In addition,
optimal two-state ensembles must consist of either two states on the $z$ axis of
the Bloch sphere, or two states on a line that is perpendicular to and
intersecting the $z$ axis. For cases where $A\notin(0,1/2)$, we have presented a
simple method to determine which form the optimal ensemble takes. This result
also enables us to determine if the input states should be orthogonal or
non-orthogonal. Even in cases where $A\in(0,1/2)$, if two states are sufficient
for the optimal ensemble, then the ensemble must take one of these two forms.

For cases where three states are necessary for the optimal ensemble, our results
show that the optimal three-state ensemble consists of one state on the $z$
axis at the maximum distance from the origin, and two states on a line
perpendicular to and intersecting the $z$ axis. This demonstrates that the form
of the optimal three state ensembles found in Ref.\ \cite{kingnath} is
universal.

Lastly, we have provided a computationally efficient method of determining the
Holevo capacity. For cases where the optimal ensemble consists of two states on
the $z$ axis, the capacity may be determined analytically. For other cases the
calculation is a numerical maximisation of a function of a single real variable,
which is easily performed. For the specific case of extremal channels, this
method is equivalent to that given in Ref.\ \cite{uhlmann}.

\acknowledgments
This project has been supported by the Australian Research Council and the
University of Queensland. The author is grateful for helpful comments from Barry
Sanders.


\begin{thebibliography}{}
\bibitem{holevo} A. S. Holevo, IEEE Trans. Info. Theory {\bf 44}, 269 (1998).
\bibitem{schuwes} B. Schumacher and M. D. Westmoreland, \pra {\bf 56}, 131
(1997).
\bibitem{davies} E. B. Davies, IEEE Trans. Info. Theory {\bf IT-24}, 596 (1978).
\bibitem{kingrus} C. King and M. B. Ruskai, IEEE Trans. Info. Theory {\bf 47},
192 (2001).
\bibitem{fuchs} C. Fuchs, \prl {\bf 79}, 1162 (1997).
\bibitem{kingnath} C. King, M. Nathanson, and M. B. Ruskai, \prl {\bf 88},
057901 (2002).
\bibitem{hayashi} M. Hayashi, H. Imai, K. Matsumoto, M. B. Ruskai, and
T. Shimono, quant-ph/0403176 (2004).
\bibitem{cortese} J. Cortese, quant-ph/0207128 (2002).
\bibitem{verst} F. Verstraete and H. Verschelde, quant-ph/0202124 (2002).
\bibitem{schu} B. Schumacher and M. D. Westmoreland, \pra {\bf 63}, 022308
(2001).
\bibitem{ohya} M. Ohya, D. Petz, and N. Watanabe, Prob. Math. Stats. {\bf 17},
170 (1997).
\bibitem{niu} C.-S. Niu and R. B. Griffiths, \pra {\bf 60}, 2764 (1999).
\bibitem{ruskai} M. B. Ruskai, S. Szarek, and E. Werner, Lin. Alg. Appl.
{\bf 347}, 159 (2002).
\bibitem{uhlmann} F. Uhlmann, J. Phys. A: Math. Gen. {\bf 34}, 7047 (2001).
\bibitem{hadamard} J. Hadamard, J. Math. Pures Appl. {\bf 58}, 171 (1893);
D. S. Mitrinovi\'c and I. B. Lackovi\'c, Aequationes Math. {\bf 28}, 229 (1985).
\end{thebibliography}
\end{document}